\documentclass[%
 reprint,
superscriptaddress,
 amsmath,amssymb,
 aps,
]{revtex4-2}

\usepackage{graphicx}% Include figure files
\usepackage{dcolumn}% Align table columns on decimal point
\usepackage{bm}% bold math
\usepackage{xcolor}
\usepackage{caption}
\usepackage{subcaption}
\begin{document}

\preprint{APS/123-QED}

\title{%The Quantum Nose \textcolor{red}{have you checked who else use the term quantum nose, I think it is too general and used to often for other issues as well https://www.nature.com/articles/s41598-017-18346-2}: Concept for a gas sensor based on quantum fluctuations \textcolor{red}{Alternative title: A novel 
A novel gas sensing principle based on quantum fluctuations}
% Force line breaks with \\
\author{Eivind Kristen Osestad}%
\affiliation{Department of Physics and Technology, University of Bergen, All\'egaten 55, 5007 Bergen, Norway.}%

\author{Pekka Parviainen}
\affiliation{Department of Informatics, University of Bergen, Thormøhlens gate 55, 5006 Bergen, Norway.}%

%\author{Bodil Holst}
%\affiliation{Department of Physics and Technology, University of Bergen, All\'egaten 55, 5007 Bergen, Norway.}
\author{Johannes Fiedler}
\email{johannes.fiedler@uib.no}
\affiliation{Department of Physics and Technology, University of Bergen, All\'egaten 55, 5007 Bergen, Norway.}%

\date{\today}% It is always \today, today,
             %  but any date may be explicitly specified

\begin{abstract}
We present a model of a novel measurement scheme to detect small amounts of a gas species via the ground-state fluctuations of the electromagnetic field (dispersion forces) depending on the entire spectral properties of all objects.

Here, we describe an experimental setup of optically trapped nanoparticles in a hollow-core fibre. We calculate the effects of the gases on the thermal motion of the nanoparticles and present a neural network-based method for reconstructing the gas concentrations. We present an example of one possible setup capable of detecting concentrations of CO2 down to 0.01 volume per cent with an accuracy of 1 ppm. 

Reliable detection of small concentrations of specific molecules in a gas is essential for numerous applications such as security and environmental monitoring, medical tests, and production processes. Unlike other measurement schemes, such as surface plasmons or functionalised surfaces, this allows for fast, continuous monitoring and using small sample quantities, without influencing the probe or the sensor system.

\end{abstract}

%\keywords{Suggested keywords}%Use showkeys class option if keyword
                              %display desired
\maketitle

%\tableofcontents

\section{Introduction}
The detection of few or even single molecules in a gas mixture is a challenge %in recent developments 
in chemical and bio-sensing~\cite{KHALILIAN2017700,Vo-Dinh2000,Yurt2012,doi:10.1021/acs.nanolett.0c04702} that has a wide range of applications, for instance, in medicine~\cite{C7TB01557G,s8052932,OH20081161} and biological research~\cite{doi:10.1126/science.1162986,Rissin2010}. One established concept uses chemical markers bound at the investigated molecule, enhancing the fluorescence~\cite{10.1093/protein/gzp074}. A similar concept works with functionalised surfaces where the molecule sticks to and chanced properties of the cantilever can be detected. In such systems, one needs to ensure that no other molecule reacts with the marker. An alternative method measures the perturbation of surface plasmons caused by the presence of dielectric molecules~\cite{Yu2021,doi:10.1021/acssensors.7b00382}. Such devices consist of a dielectric body carrying a surface plasmon. When a molecule approaches such a device, the surface plasmon resonance is spectrally detuned due to the field enhancements caused by the molecule's presence, which should be detected. %Due to the applied dipole coupling, the perturbing dipole moment has to be evaluated at the specific frequency ($\omega_0$) of the resonance~${\bm{d}} = \alpha(\omega_0)\cdot {\bm{E}}(\omega_0)$, which is not a unique quantity for all molecules. 
To this end, two issues can be identified with the existing concepts for single-molecule measurements: (i) selectivity; to achieve it for few- or single-molecule measurements, the probe is usually destroyed afterwards, and (ii) additional knowledge about the constituents of the probe is required. In contrast, by investigating macroscopic objects, the electromagnetic spectrum is used to identify a specific molecule, for instance, in chromatography~\cite{GONZALEZGONZALEZ2020105}. With the proposed method, %havent't proposed the method yet 
we want to transfer the concept of identifying single molecules via their electromagnetic response measurements to the microscopic scale to detect single- or few-molecule ingredients.

\begin{figure}[t]
   % \centering
    \includegraphics[width=\columnwidth]{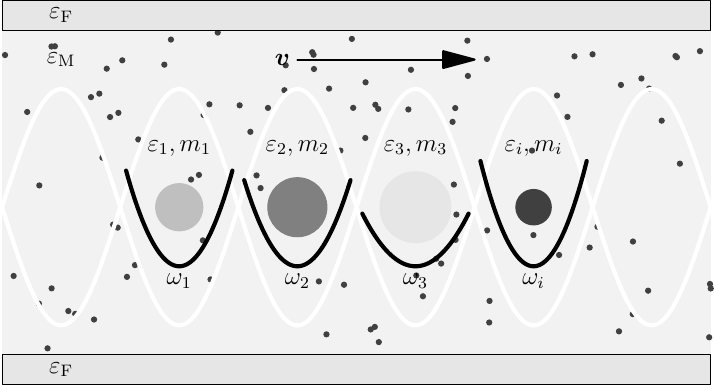}
    \caption{Schematic illustration of the considered scenario: a hollow core fibre with permittivity $\varepsilon_{\rm F}$ is doped with optically trapped nanoparticles with dielectric functions $\varepsilon_i$ and masses $m_i$ within a standing laser field (white line). A constant, homogeneous flow of a gas mixture (grey dots) with velocity ${\bm{v}}$ and bulk dielectric function $\varepsilon_{\rm M}$ passes the nanoparticles through the fibre. It perturbs the trapping potential due to the screening of the fields.}
    \label{fig:scheme}
\end{figure}

We adapt the experiment reported in Refs.~\cite{10.3389/fbioe.2021.637715,Epple2014} by adding a gas mixture and demonstrating the impact on the experimental measurements due to the modified interactions. We consider a series of optically trapped nanospheres with different masses and different optical responses each, as illustrated in Fig.~\ref{fig:scheme}. These spheres are embedded in a hollow-core fibre and stabilised due to induced dipole interaction. Experiments with a similar setup have already been performed~\cite{Epple2014}. Due to the thermal fluctuations, the particles will automatically oscillate within the trapping potential. A gas flowing through such fibres will, on the one hand, increase the elongation of the oscillations and, on the other hand, perturb the trapping potential due to the screening of the induced dipole forces. Of course, the nanoparticles can be kicked out of the trap for gas flows with high momentum. However, we restrict ourselves to the considerations of gas flows with low velocity such that the harmonic oscillation approach will still be valid. A specific molecular dynamics simulation will be the content of further investigations.

In previous work, it has been demonstrated that a series of dispersion force measurements can determine the dielectric function of a material. In particular, the measurements of Hamaker constants between two plates embedded in a two-component liquid have been investigated. It has been shown that the dynamic dielectric response of one component, either one of the plates or one of the liquids, is uniquely mapped onto the Hamaker constants with different liquid concentrations~\cite{PhysRevApplied.13.014025}. 

This uniqueness of the Hamaker constants for continuous concentrations and the dielectric response function will be transferred onto the Casimir--Polder interaction between the nanospheres and the hollow-core fibre.%, leading to a mapping issue concerning a finite number of measurements.
%Here, we present a novel measurement scheme consisting of a series of optically trapped nanoparticles embedded in a hollow-core fibre with a homogeneous gas mixture flowing homogeneously through. 
We introduce the theoretical model describing the dependency of the trapping frequencies on the gas composition. The result is a non-linear system of equations mapping the concentrations expressed by the partial pressure of all gas components on the trapping frequencies. To identify the concentrations of a single component, we will further present an inversion method via machine learning and apply this method on a gas mixture consisting of ten few-atomic molecules (CO$_2$, CH$_4$, N$_2$O, O$_2$, O$_3$, NO, CO, NO$_2$, H$_2$S, and N$_2$) to identify the concentration of CO$_2$.

\section{Theoretical Modelling}
To describe the relationship between the trapping frequencies and the gas mixture, we concentrate on the two most dominant interactions for optically trapped neutral particles within a neutral environment: the optical potential~\cite{Fiedler_2017}
\begin{equation}
    U_{\rm EM}({\bm{r}}_{\rm A}) = -\frac{1}{2}\alpha(\lambda) {\bm{E}}^2 ({\bm{r}}_{\rm A},\lambda) \,,\label{eq:Uopt}
\end{equation}
with the laser frequency $\lambda$ and the particle's position and optical response, ${\bm{r}}_{\rm A}$ and $\alpha$, respectively, due to the laser field which traps the nanoparticles, in general, and the Casimir--Polder potential~\cite{Scheel2008,Buhmann12a,Buhmann12b}
\begin{equation}
   U_{\rm CP}({\bm{r}}_{\rm A}) = \mu_0 k_{\rm B}T \sum_{n=0}^\infty {}'\xi^2 \operatorname{Tr} \left[ {\bm{\alpha}}({\rm i}\xi_n)\cdot {\bf{G}}({\bm{r}}_{\rm A},{\bm{r}}_{\rm A},{\rm i}\xi_n)\right] \,,\label{eq:UCP}
\end{equation}
with the vacuum permeability $\mu_0$, the Boltzmann constant $k_{\rm B}$, the temperature $T$, and the scattering Green function ${\bf{G}}$. Due to the finite temperature, the optical modes are restricted to the Matsubara frequencies $\xi_n = (2\pi k_{\rm B} T/\hbar) n$ with an integer $n$. The primed sign at the sum denotes that the zeroth term has to be weighted by $1/2$. The Casimir--Polder interaction~(\ref{eq:UCP}) can be understood as an exchange of virtual photons ${\rm i}\xi_n$. Caused by the ground-state fluctuation of the electromagnetic field, the particle at position ${\bm{r}}_{\rm A}$ gets excited and emits virtual photons that are scattered back by the environment, ${\bf{G}}({\bm{r}}_{\rm A},{\bm{r}}_{\rm A},{\rm i}\xi_n)$~\cite{doi:10.1142/9383}. As virtual photons obey the same propagation rules as real photons, they can be diffracted at interfaces, leading, for instance, to the dispersion forces, and absorbed by media~\cite{doi:10.1063/5.0037629}. Further interactions neglected within this study are induced forces due to the presence of the laser field~\cite{PhysRevA.98.022514} and field modifications due to the reflection at and the penetration into the fibre of the laser field.

We assume the nanoparticles with radius $a$ to be small compared to the hollow-core inner radius $R_{\rm i}$. This allows us to determine its optical response via a Clausius--Mossotti-like relation~\cite{2011661,Jackson}, which is known as the hard-sphere model for access polarisabilities~\cite{doi:10.1021/acs.jpca.7b10159}
\begin{equation}
    \alpha^\star (\omega) = 4\pi \varepsilon_0\varepsilon_{\rm M}(\omega) a^3 \frac{\varepsilon_{\rm NP}(\omega) -\varepsilon_{\rm M}(\omega)}{\varepsilon_{\rm NP}(\omega) + 2\varepsilon_{\rm M}(\omega)}\,, \label{eq:HS}
\end{equation}
with the nanoparticle's dielectric function $\varepsilon_{\rm NP}$.

To consider the curvature of the hollow core fibre, we apply the local-field corrected Born series expansion for the scattering Green function~\cite{Scheel2008,Buhmann12b}. This approach separates the scattering processes according to the number of scattering events. Its first order reads~\cite{PhysRevX.4.011029}
\begin{eqnarray}
    \lefteqn{{\bf{G}}({\bm{r}}_{\rm A},{\bm{r}}_{\rm A},\omega) = \frac{\omega^2}{c^2}\int\limits_{\rm V} \frac{\mathrm d^3 s\chi(\omega)}{1+\chi(\omega)/3}}\nonumber\\
    &&\times{\bf{R}}({\bm{r}}_{\rm A},{\bm{s}},\omega)\cdot {\bf{R}}({\bm{s}},{\bm{r}}_{\rm A},\omega)\,,\label{eq:Green}
\end{eqnarray}
with the susceptibility of the fibre relative to the medium $\chi(\omega) = \varepsilon_{\rm F}(\omega) -\varepsilon_{\rm M}(\omega)$ and the regular part of the bulk Green function~\cite{Scheel2008}
\begin{eqnarray}
    {\bf{R}}({\bm{r}},{\bm{r}}',\omega)  = \frac{q}{4\pi }\left[f\left(\frac{1}{q\varrho}\right)\mathbb{I}- g\left(\frac{1}{q\varrho}\right)\frac{{\bm{\varrho}}{\bm{\varrho}}}{\varrho^2}\right]\mathrm e^{{\rm i}q \varrho },
\end{eqnarray}
with the three-dimensional unit matrix $\mathbb{I}$, the relative coordinate ${\bm{\varrho}}={\bm{r}}-{\bm{r}}'$, its absolute value $\varrho=\left|{\bm{\varrho}}\right|$, absolute value of the wave vector $q= \sqrt{\varepsilon_{\rm M}}\omega/c$ and the functions $f(x) = x+{\rm i}x^2-x^3$ and $g(x) = x+3{\rm i}x^2-3x^3$. Hence, the scattering Green function~(\ref{eq:Green}) is constructed via the single scattering events with all volume elements $\mathrm d^3 s$ of the fibre. Thus, the integral volume is given by the volume of the hollow-core fibre, $\int_{\rm V} \mathrm d^3 s = \int_{-\infty}^\infty \mathrm d z_s \int_0^{2\pi}\mathrm d \varphi_s \int_{R_{\rm i}}^{R_{\rm o}}\mathrm d r_s\,r_s$.

To model the impact of the gas mixture components, we assume a simple Lorentz-oscillator model for their single optical responses
\begin{eqnarray}
    \alpha_i({\rm i}\xi) = \sum_j \frac{c_{i,j}}{1+(\xi/\omega_{i,j})^2}\,,
\end{eqnarray}
with a set of resonance frequencies $\omega_{i,j}$ and oscillator strengths $c_{i,j}$ for each component $i$. By assuming further that the gas mixture behaves like an ideal gas $N/V=P/(k_{\rm B}T)$, the polarisability per volume is given by
\begin{equation}
    \tilde{\alpha}_i({\rm i}\xi ) = \frac{P_i\alpha_i({\rm i}\xi)}{k_{\rm B}T} \,,
\end{equation}
with the partial pressure of the $i$th component $P_i$. Thus, the total polarisability of the gas mixture can be obtained by summing over components
\begin{eqnarray}
    \alpha_{\rm mix}({\rm i}\xi) = \sum_i \tilde{\alpha}_i({\rm i}\xi) \,.
\end{eqnarray}
By applying effective medium theory, this polarisability can be transformed into an effective permittivity for the gas mixture~\cite{Aspnes}
\begin{equation}
    \varepsilon_{\rm M}({\rm i}\xi ) = \frac{1+ 2\alpha_{\rm mix}({\rm i}\xi)}{1-\alpha_{\rm mix}({\rm i}\xi)}\,.\label{eq:mix}
\end{equation}

By combining the optical potential~(\ref{eq:Uopt}) with the Casimir--Polder potential~(\ref{eq:UCP}), the total interaction potential can be obtained, leading to the axial trapping frequency
\begin{eqnarray}
    \omega_z = \sqrt{\frac{\partial_z^2 U_{\rm EM}({\bm{0}})}{m}} \,, \label{eq:omez}
\end{eqnarray}
and the radial trapping frequency
\begin{eqnarray}
    \omega_r = \sqrt{\frac{\partial_r^2 U_{\rm EM}({\bm{0}})+\partial_r^2 U_{\rm CP}({\bm{0}})}{m}} \,, \label{eq:omer}
\end{eqnarray}
where we approximated the real-valued optical response for the laser field according to its imaginary $\varepsilon(\omega) \approx \varepsilon({\rm i}\omega)$, which is a suitable approximation for frequencies in visible range~\cite{doi:10.1063/5.0037629}, for the evaluation of the optical potential which is modelled via a standing laser field
\begin{equation}
    {\bm{E}}({\bm{r}}) = {\bm{E}}_0 \mathrm e^{-2r^2/R^2} \cos\left(\frac{\overline{\omega} z}{c}\right)\,,
\end{equation}
with the laser radius $R$ and frequency $\overline{\omega}$. This model neglects the actual propagation of the laser field through the fibre. It assumes that the laser beam is small enough not to affect the fibre's walls. For wider beams, parts of the beam will be reflected from the walls and thus enhance the fields along the centre line. Together with the gas mixing model~(\ref{eq:mix}), the impact of changing the gas composition on the trapping frequency~(\ref{eq:omez}) and (\ref{eq:omer}) is obtainable.

\begin{figure}[t]
    \centering
    \includegraphics[width=0.8\columnwidth]{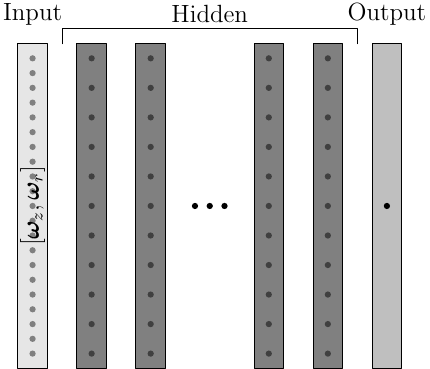}
    \caption{Overview of the neural network architecture: the input layer getting the oscillator frequencies is fully connected with the 12 hidden layers with 288 nodes each, which produces the single value output (CO$_2$ concentration). All layers are activated with a ReLU unit.}
    \label{fig:ANN}
\end{figure}

\section{Neural Network}
We consider a sensor as depicted in Fig.~\ref{fig:scheme} with ten spheres. We receive 20 unique trapping frequencies for any gas mixture according to Eqs.~\eqref{eq:omer} and \eqref{eq:omez}. We use the Keras Deep learning library~\cite{chollet2015keras} to create a densely connected neural network architecture as illustrated in Fig.~\ref{fig:ANN}. The input is a vector $\hat{\bm{\omega}}=[{\bm{\omega}}_z,{\bm{\omega}}_r]$ with 20 trapping frequencies, but we normalise component wise 
\begin{eqnarray}
    \hat{\omega}_{\rm i} = \frac{\omega_{\rm i} - \min \omega_{\rm i}}{\max \omega_{\rm i} - \min \omega_{\rm i}}\,,
\end{eqnarray}
where $\min \omega_{\rm i}$ is the minimum value of a trapping frequency in the data set, and $\max \omega_{\rm i}$ is the maximum value. This way, all inputs are between 0 and 1. We then have 12 hidden, fully connected dense layers with 288 units. The $\lambda$-th dense hidden layers are given by the previously hidden layer by  
\begin{eqnarray}
    \label{eq:layer}
    {x}^{(\lambda)}_i = \sum_{j} {W}_{ij}^{(\lambda)} {x}^{(\lambda-1)}_j + {b}_i^{(\lambda)}\,,
\end{eqnarray}
where ${W}_{ij}^{(\lambda)}$ and ${b}_i^{(\lambda)}$ are trainable parameters. 
We apply the Rectified Linear Unit (ReLU) activation to each hidden layer, which takes a value $x$ and gives $\max\{0, x\}$. The output layer is densely connected with a single unit and ReLU activation. An overview of the network is shown in Fig.~\ref{fig:ANN}.

\begin{figure}[t]
    \centering
    \includegraphics[width=\columnwidth]{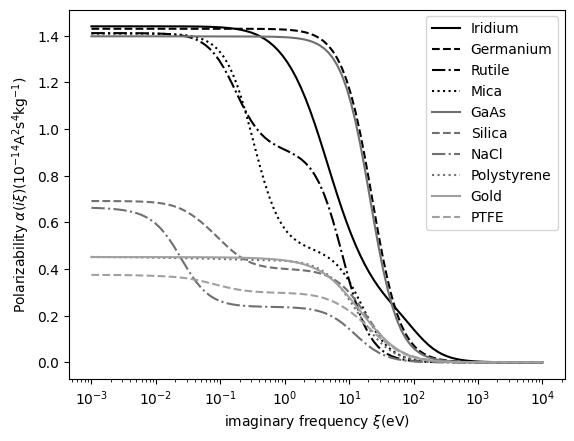}
    \caption{Polarisablities of nanoparticles with 10~nm radius for Silica, PTFE, polystyrene, mica, rutile, gold, gallium arsenide, germanium, sodium chloride and iridium.}
    \label{fig:nanoalpha}
\end{figure}

The training was done using RMSprop~\cite{RMSprop}. We start with a learning rate of $10^{-7}$. Then, we train for 500 epochs. After this, we lower the learning rate to $10^{-8}$ for 500 epochs. Then, we do 500 epochs at a learning rate of $10^{-9}$ and then at $10^{-10}$. In total, we train for 2000 epochs. For our loss function, we use mean square error. We use a batch size of 100.

\begin{figure}[t]
    \centering
    \includegraphics[width=\columnwidth]{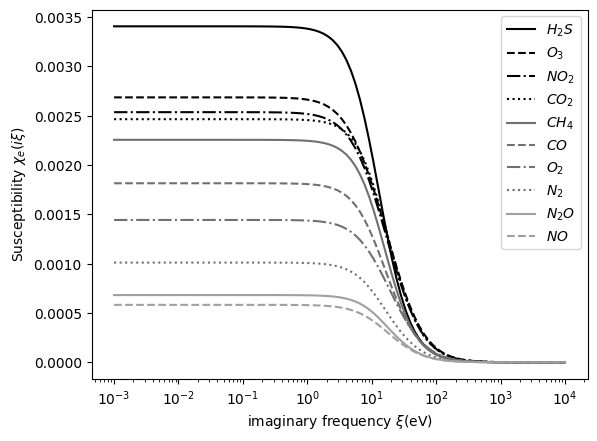}
    \caption{Dielectric susceptibilities of the gas mixture constituents: carbon dioxide (CO$_2$), methane (CH$_4$), nitrous oxide (N$_2$O), ozone (O$_3$), oxygen (O$_2$), nitric oxide (NO), carbon monoxide (CO), nitrogen dioxide (NO$_2$), hydrogen sulfide (H$_2$S), and nitrogen (N$_2$).}
    \label{fig:gaschi}
\end{figure}

\section{Measuring the CO$_2$ concentration within a gas mixture}
We considered the gas sensor to be built off silicon dioxide (SiO$_2$) with an inner radius $R_{\rm i}= 500\,\rm{nm}$ and an outer radius $R_{\rm o} = 1000\,\rm{nm}$. We applied the two-oscillator model from Ref.~\cite{Hemmerich2016} to model its dielectric response. Furthermore, we considered the following materials for the dielectric nanoparticles: silica, polytetrafluoroethylene (PTFE), and polystyrene (oscillator models are reported in Ref.~\cite{PhysRevA.81.062502}); mica, rutile, and gold (oscillator models are reported in Ref.~\cite{doi:10.1021/acs.jpcc.0c06748}); gallium arsenide, germanium, sodium chloride and iridium (measured values are taken from Ref.~\cite{palik1998handbook}, and by applying Kramers--Kronig relation~\cite{Jackson}, they have been transferred to the imaginary frequency axis). The resulting vacuum polarisabilities are depicted in Fig.~\ref{fig:nanoalpha}, where we applied the Clausius--Mossotti relation (Eq.~\eqref{eq:HS} with $\varepsilon_{\rm M}  = 1$) together with the particle radius of 10~nm. It can be seen that all response functions strongly differ from each other due to the different resonances of the materials. Furthermore, gallium arsenide and germanium are very similar. The dynamical response functions for gold and polystyrene are almost identical within a broad spectral range. However, the Debye contribution for gold at lower frequencies enhances their distinguishability.
\begin{figure*}[t]
    \centering
        \begin{subfigure}[t]{0.49\textwidth}
        \centering
        \includegraphics[width=\linewidth]{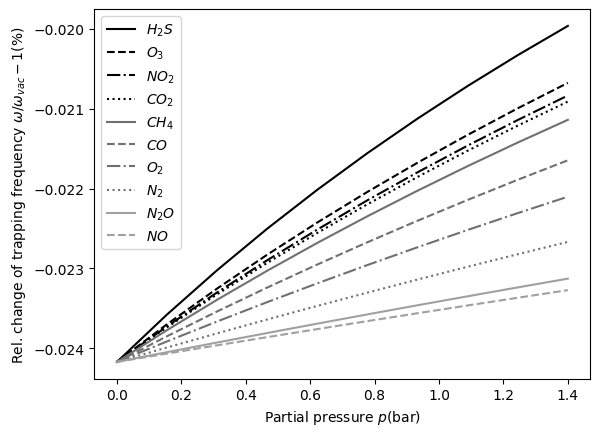} 
        \caption{Silica}
    \end{subfigure}
    \hfill
    \begin{subfigure}[t]{0.49\textwidth}
        \centering
        \includegraphics[width=\linewidth]{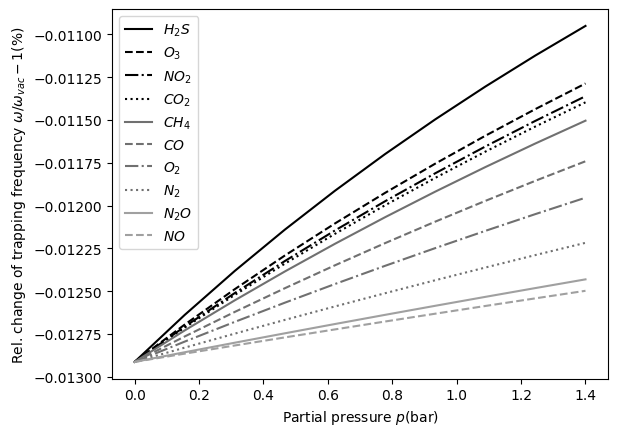} 
        \caption{Gold} 
    \end{subfigure}

    \vspace{0.2cm}
    \begin{subfigure}[t]{0.49\textwidth}
        \centering
        \includegraphics[width=\linewidth]{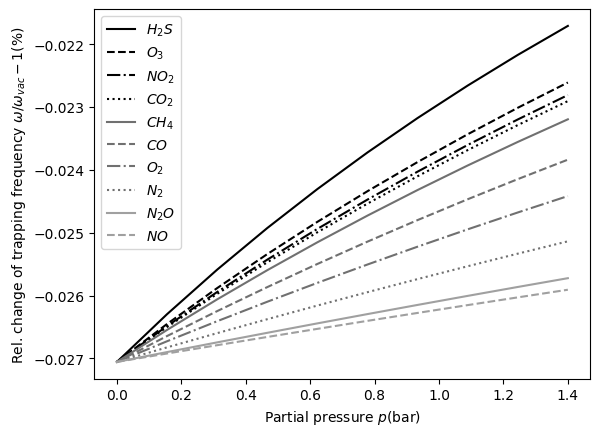} 
        \caption{NaCl}
    \end{subfigure}
    \hfill
    \begin{subfigure}[t]{0.49\textwidth}
        \centering
        \includegraphics[width=\linewidth]{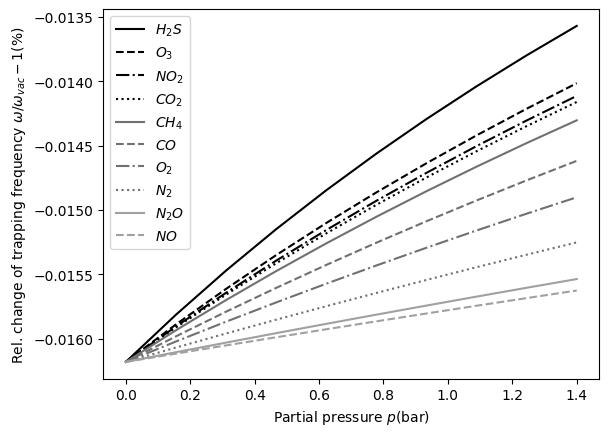} 
        \caption{PTFE} 
    \end{subfigure}
    \caption{Relative change of the radial trapping frequency~(\ref{eq:omer}) for (a) silica, (b) gold, (c) sodium chloride (NaCl) and (d) PTFE nanoparticles.}
    \label{fig:trapfreq}
\end{figure*}
We considered the gas sensor to be built off silicon dioxide (SiO$_2$) with an inner radius $R_{\rm i}= 500\,\rm{nm}$ and an outer radius $R_{\rm o} = 1000\,\rm{nm}$. We applied the two-oscillator model from Ref.~\cite{Hemmerich2016} to model its dielectric response. Furthermore, we considered the following materials for the dielectric nanoparticles: silica, polytetrafluoroethylene (PTFE), and polystyrene (oscillator models are reported in Ref.~\cite{PhysRevA.81.062502}); mica, rutile, and gold (oscillator models are reported in Ref.~\cite{doi:10.1021/acs.jpcc.0c06748}); gallium arsenide, germanium, sodium chloride and iridium (measured values are taken from Ref.~\cite{palik1998handbook}, and by applying Kramers--Kronig relation~\cite{Jackson}, they have been transferred to the imaginary frequency axis). The resulting vacuum polarisabilities are depicted in Fig.~\ref{fig:nanoalpha}, where we applied the Clausius--Mossotti relation (Eq.~\eqref{eq:HS} with $\varepsilon_{\rm M}  = 1$) together with the particle radius of 10~nm. It can be seen that all response functions strongly differ from each other due to the different resonances of the materials. Furthermore, gallium arsenide and germanium are very similar. The dynamical response functions for gold and polystyrene are almost identical within a broad spectral range. However, the Debye contribution for gold at lower frequencies enhances their distinguishability.

The gas mixture that we considered within this simulation was made out of the following few-atomic molecules: carbon dioxide (CO$_2$), methane (CH$_4$), nitrous oxide (N$_2$O), ozone (O$_3$), oxygen (O$_2$), nitric oxide (NO), carbon monoxide (CO), nitrogen dioxide (NO$_2$), hydrogen sulfide (H$_2$S), and nitrogen (N$_2$) with randomly assigned concentrations with a partial pressure between 0 and 0.2 bar. The corresponding polarisabilities and molecular volumes have been taken from Refs.~\cite{doi:10.1021/acs.jpca.7b10159,C9CP03165K}. The resulting dielectric susceptibilities $\chi=\varepsilon-1$ according to the mixing model~(\ref{eq:mix}) are depicted in Fig.~\ref{fig:gaschi}. It can be observed that the optical transitions dominate the dielectric responses because all functions drop down to zero in the optical range. Thus, the main criterion for their reconstruction is the magnitude of the response. 

We considered an intensely focused laser beam with a beam width of $2R=986.5\,\rm{nm}$ to avoid field enhancements due to the interaction with the fibre, a laser power of $0.0015\,\rm{nW}$ at a wavelength of $\lambda=895\,\rm{nm}$. Such a laser system is experimentally achievable~\cite{Epple2014} and sufficient to trap the nanoparticles stably. The resulting impact of changing the partial pressure of the gas components on the radial trapping frequency~(\ref{eq:omer}) are depicted in Fig.~\ref{fig:trapfreq}, exemplary for silica, gold, sodium chloride and PTFE. It can be observed that the trapping frequency increases with the pressure (equivalent to the particle density) in general. The most interesting behaviour shows H$_2$S and O$_3$ with a PTFE nanoparticle, which roughly takes a maximal frequency shift at standard pressure for H$_2$S and 1.2 bar for O$_3$. After that, the change of the trapping frequency decreases with increasing pressure. This behaviour illustrates the nature of this method. According to the Casimir--Polder potential~\eqref{eq:UCP} together with the polarisability model~\eqref{eq:HS}, each trapping frequency is a comparison between the dielectric function of the sphere, Fig.~\ref{fig:nanoalpha}, with the dielectric function of the gas mixture, Fig.~\ref{fig:gaschi}, which can result in repulsive forces~\cite{PhysRevB.101.235424}. Several oscillators are required as each trapping frequency provides a single measure for this dielectric function comparison.

We generated two data sets for training the densely connected neural network. One was generated by picking nine random numbers $a_i$ uniformly between 0 and 0.2 bar with summation restriction $\sum_i a_i = 0.2\,\rm{bar}$. We restricted the total number of particles in the gas to this pressure to reduce the impact of their Brownian motion~\cite{doi:10.1126/science.abg3027}.
According to this partial pressure distribution, the resulting trapping frequencies ${\bm{\omega}}_z$ and ${\bm{\omega}}_r$ have been calculated for all spheres. The second set is generated similarly, but the CO$_2$ concentration is between 0 bar and 0.01 bar. We generated 100,000 gas mixtures each for both sets. 

\begin{figure}[t]
    \centering
    \includegraphics[width=\columnwidth]{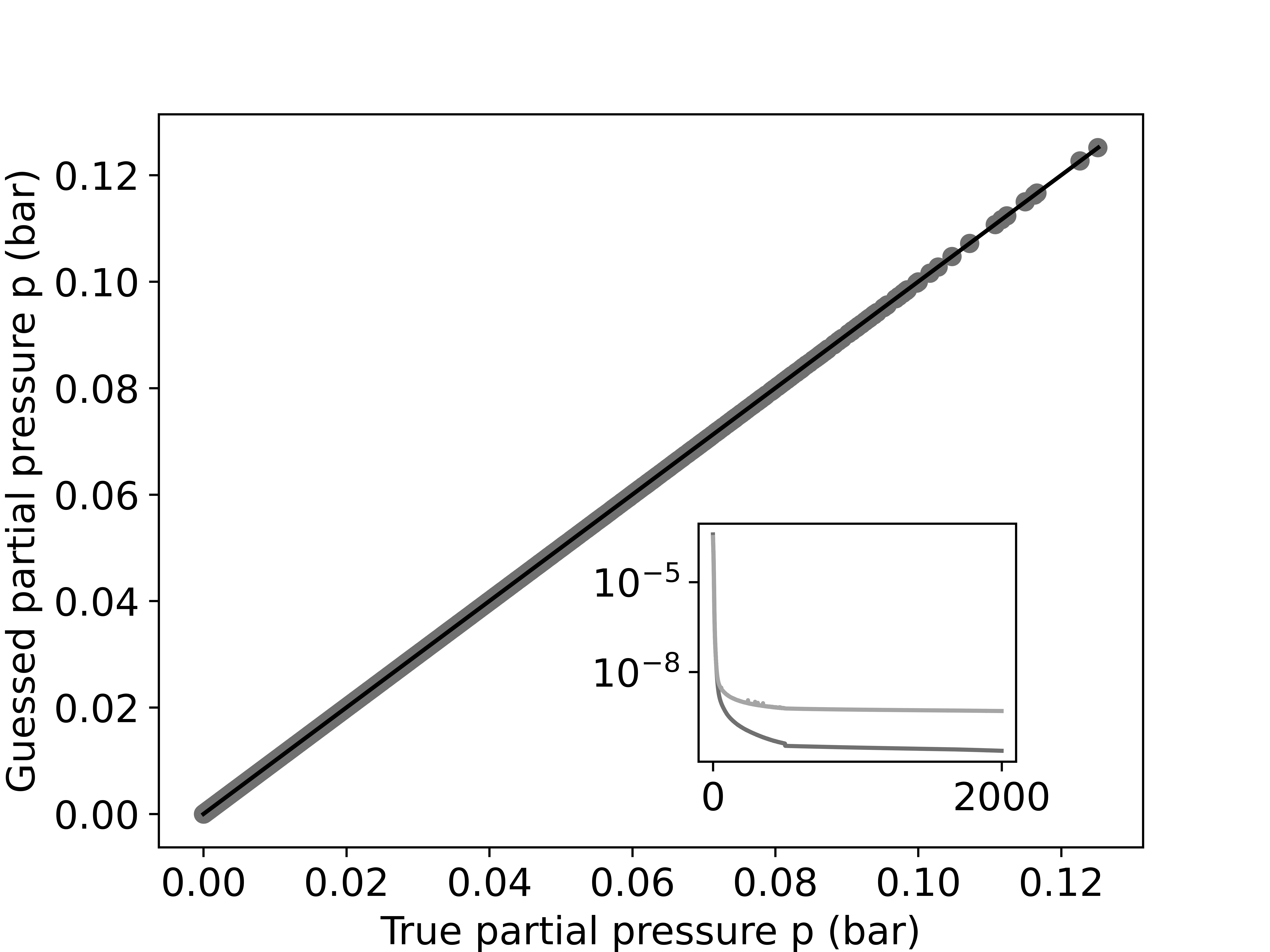}
    \caption{True value of CO$_2$ concentration against the first neural network guesses for the validation data (grey scatter plot). The black line shows the line where perfect guesses would land. The inset shows the training loss (grey lower) and validation loss (light grey) during training.}
    \label{fig:perf}
\end{figure}

Using the abovementioned method, We trained one neural network on each data set. We picked 20,000 gas mixtures from each data set to be used as validation data. The final guesses plotted against the true values of CO$_2$ concentration for the first 10000 data points in the first data set validation data are shown in Fig.~\ref{fig:perf}. We see that the guesses and true values are very close. The final validation loss was 4.97$\times 10^{-10}$ bar$^2$. As our loss function is a mean square error, the square root of this is an approximation of the sensor's uncertainty. The square root of the validation loss of the first network yields an uncertainty of $\Delta p = 2.23\times 10^{-5}$ bar. We see overfitting early, but the validation loss decreases until somewhere before epoch 500, where it flattens out.  

\begin{figure}[t]
    \centering
    \includegraphics[width=\columnwidth]{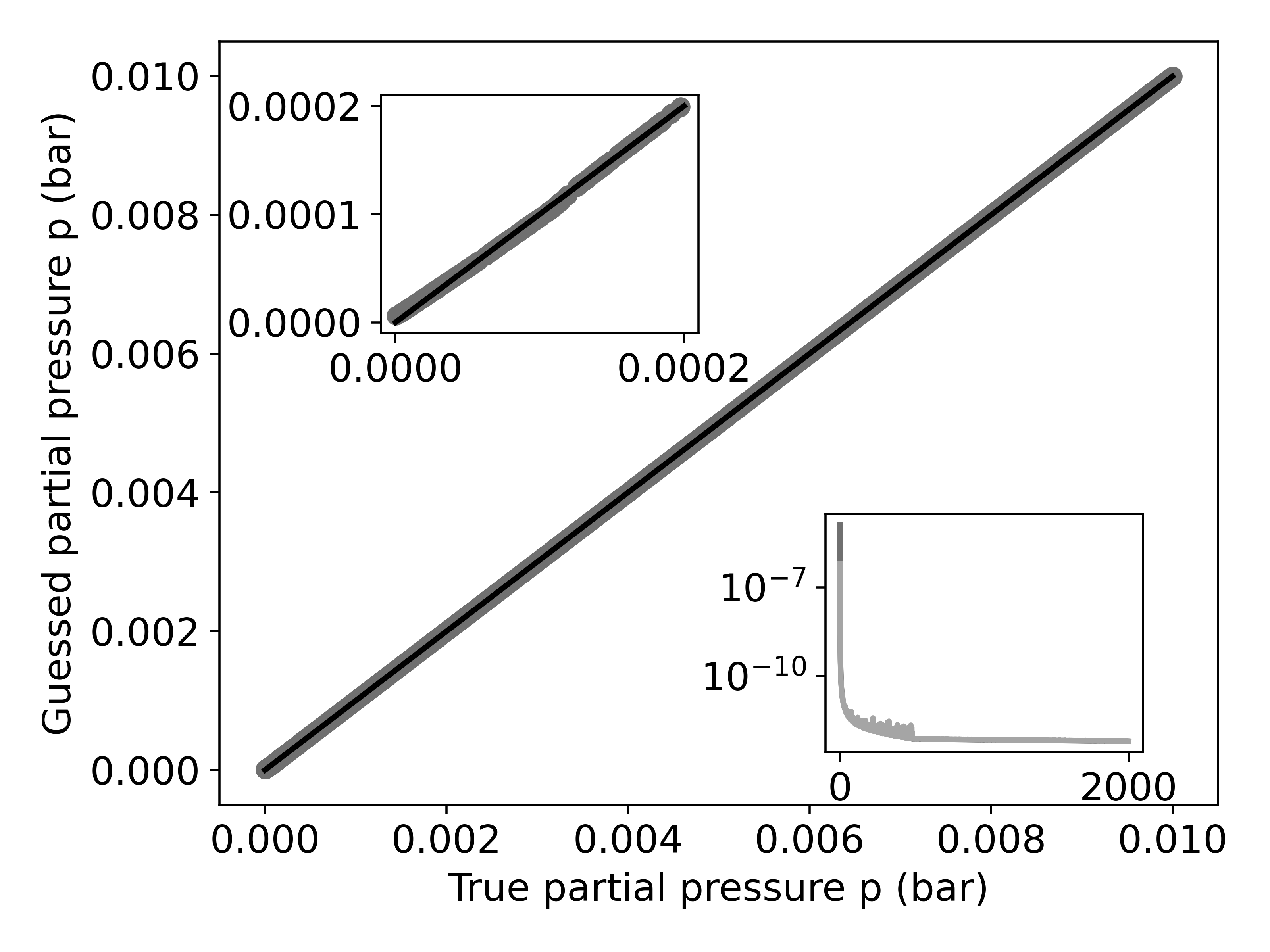}
    \caption{True value of CO$_2$ concentration against the second neural network guesses for the validation data (grey scatter plot). The black line shows the line where perfect guesses would land. The top left inset shows the guesses from 0 to 0.0002. The bottom right inset shows the training loss (grey) and validation loss (light gray) during training.}
    \label{fig:perflow}
\end{figure}

The network trained on the second data set with lower CO$_2$ concentrations has its final validation guesses on the first 10000 data points in its validation data set, shown in Fig.~\ref{fig:perflow}. Here, the final validation loss is $6.19\times10^{-13}$ bar$^2$. Its square root yields an uncertainty of $\Delta p =7.87\times10^{-7}$ bar. The validation loss and training loss were much closer during training. In the end, the training loss was $6.12\times10^{-13}$ bar$^2$. We saw no decrease in the loss by training it further.

\section{Conclusions}

We present a measurement scheme determining the gas concentration within a gas mixture. We demonstrated the functionality of such a sensor by determining the gas concentration of CO$_2$ within a ten-component mixture down to 0 volume per cent with an accuracy of around one ppm. For very low concentrations down to single-molecule detection, further investigations are required. The proposed device is based on an existing experimental setup and considers the impact of a surrounding gas via the coupling to the quantum vacuum. The proposed measurement scheme will be fast due to reduced measuring trapping frequencies.

To demonstrate the general reliability, we restricted to a single gas component. However, the simultaneous detection of several components is directly implementable, but the amount of training time grows almost exponentially with the number of gas components. Furthermore, we restricted mostly to simple two- or three-atomic gas molecules, which respond most actively in the optical and ultraviolet spectral range. Concerning the application of the measurement scheme, more complex molecules with electromagnetic responses covering a larger spectrum have to be considered, leading to an optimisation task for the materials of the spheres to be sensitive in these specific ranges. Such materials can be generated via functionalising the surfaces~\cite{D1AN00582K,doi:10.1021/acs.langmuir.1c01525,doi:10.1021/acs.jctc.9b01251,D2CP03349F}. 

A practical implementation of the proposed measurement scheme would yield imperfections of the spherical nanoparticles in shape and mass density, leading to a slightly different potential landscape trapping the particle and, hence, slightly changed trapping frequencies. Also, the laser field will be polychrome, modifying the trapping potential. Such experimental imperfections can be incorporated via an adequate learning strategy, starting with many ideal cases followed by a set of controlled measurements. Of course, even the ideal cases need to be adapted concerning the correct propagation of the laser beam within the hollow-core fibre, leading to an enhancement of the field strength and due to the particle interactions concerning induced dipole moments within the spheres and the gas molecules leading to further interaction. Furthermore, an application to liquids is possible as well but will yield several further issues. For instance, the optical trapping of nanoparticles in a medium is experimentally challenging, the consideration of a solvent typically yields further interactions due to the dissociation of the liquids leading to additional electrostatic forces, and the microscopic fluid dynamics need to be considered explicitly. However, the impact of such effects will be the content of further investigations.

\begin{acknowledgments}
We thank Bodil Holst and J\"orn Andreas Kersten for their critical discussions and for supporting this manuscript.
J.F. gratefully acknowledges support from the European Union (H2020-MSCA-IF-2020, grant number: 101031712).
\end{acknowledgments}

\appendix

\bibliography{bibi}% Produces the bibliography via BibTeX.

\end{document}